\documentclass[prb,twocolumn]{revtex4-1}

\usepackage{graphicx}
\usepackage{amsmath,amssymb,bm}
\usepackage{dcolumn}
\usepackage{booktabs}
\usepackage{subfig}
\usepackage{xcolor}
\usepackage{array}

\usepackage{tikz}
\usetikzlibrary{arrows.meta,decorations.pathmorphing}

\newcolumntype{M}[1]{>{\centering\arraybackslash}m{#1}}

\def\be{\begin{equation}}
\def\ee{\end{equation}}
\def\bearr{\begin{eqnarray}}
\def\eearr{\end{eqnarray}}

\usepackage[colorlinks=true,linkcolor=red]{hyperref}

\begin{document}
	
	\title{Lorentz-Covariant Landau Levels of Tilted Dirac Fermions in Nonuniform Fields}
	\bigskip
	\author{Sushmita Saha and Alestin Mawrie}
	\normalsize
	\affiliation{Department of Physics, Indian Institute of Technology Indore, Simrol, Indore-453552, India}
	\date{\today}
\begin{abstract}
We develop an analytical theory of Landau quantization for tilted anisotropic Dirac fermions in an exponentially decaying magnetic field. Using anisotropy scaling and a Lorentz transformation, we recast the laboratory-frame problem into the isotropic Dirac equation in the boosted frame, where the exponentially decaying magnetic-field problem admits an exact solution. Transforming the boosted-frame spectrum back to the laboratory frame yields an implicit quantization condition with an intrinsically energy-dependent guiding-center parameter. We identify the conditions for physically admissible states. We further show that the formalism recovers the known uniform-field spectrum of tilted anisotropic Dirac fermions in the appropriate limit. Our results extend the Lorentz-covariant treatment of tilted anisotropic Dirac fermions to an exponentially decaying magnetic field and reveal the energy-dependent guiding-center structure generated by the inverse transformation to the laboratory frame.
\end{abstract}

\email{amawrie@iiti.ac.in}
	
\maketitle
\begin{figure*}[t]
	\centering
	\includegraphics[width=1.0\textwidth,height=2.75cm]{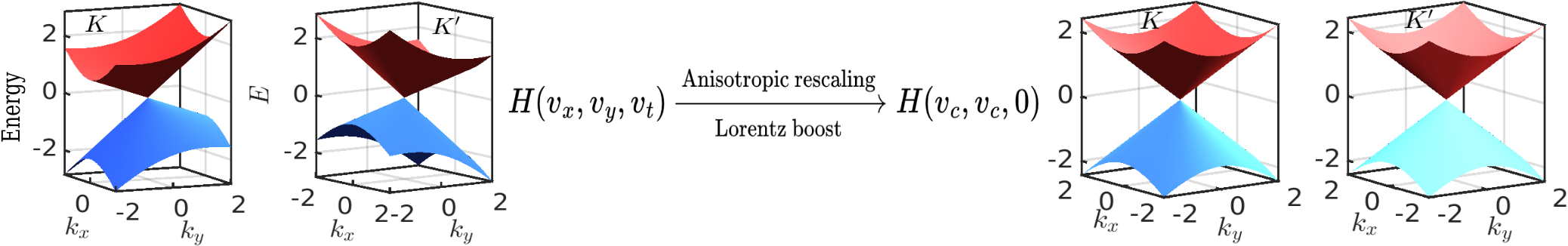}
	\caption{
		(Color online) Lorentz-covariant mapping of the tilted anisotropic Dirac
		cones in 8-\emph{Pmmn} borophene. After anisotropy rescaling and
		Lorentz boosts ($\beta=\pm v_t/v_c$), the valley-dependent tilt is
		eliminated, yielding the common isotropic Hamiltonian
		$H(v_c,v_c,0)$ with $v_c=\sqrt{v_xv_y}$.
	}
	\label{fig:borophene_cones}
\end{figure*}
\textbf{\textit{Introduction}:}
The discovery of two-dimensional Dirac materials has greatly expanded the study of relativistic quantum phenomena in condensed-matter systems.
\cite{Novoselov2005,Zhang2005,CastroNeto2009,Kotov2012,Wehling2014,Armitage2018,Bradlyn2016}
Among these, 8-\emph{Pmmn} borophene has emerged as a unique platform hosting intrinsically tilted and anisotropic Dirac cones. Its low-energy electronic structure has been established from first-principles and tight-binding studies,\cite{LopezBezanilla2016,Nakhaee2018} while strain engineering, collective excitations, optical, and linear-response properties have been extensively investigated.\cite{Zabolotskiy2016,Sadhukhan2017,Verma2017,Mandal2025,SharmaMandal2026} These band-structure characteristics give rise to unconventional transport, optical, and magnetotransport phenomena.\cite{Goerbig2008,Trescher2015,Sadhukhan2017,Verma2017} Following theoretical predictions, atomically thin borophene was successfully synthesized, stimulating extensive theoretical and experimental studies of its electronic properties.\cite{Zhou2014,Mannix2015,Feng2016}

The corresponding low-energy quasiparticles are governed by the effective Hamiltonian
\(
H=v_x\sigma_xp_x+v_y\sigma_yp_y+v_tp_y\sigma_0,
\)
where the anisotropy is encoded in the Fermi velocities \(v_x\) and \(v_y\), while \(v_t\) describes the intrinsic tilt of the Dirac cone.
A key feature of this Hamiltonian is its emergent Lorentz covariance. After anisotropy scaling, the tilt is removed exactly by a Lorentz boost, mapping the system onto an untilted isotropic Dirac Hamiltonian.\cite{Lukose2007,Goerbig2011,Islam2017,Farajollahpour2019} Time-reversal symmetry ensures opposite tilt directions in the two inequivalent valleys, which therefore require opposite boosts but map onto the same boosted Hamiltonian (Fig.~\ref{fig:borophene_cones}). This emergent symmetry underpins exact analytical descriptions of Landau quantization and magnetotransport in tilted Dirac materials under uniform magnetic fields.\cite{Islam2017,Islam2018}

Spatially nonuniform magnetic fields (NUMFs) provide an additional route for manipulating Dirac quasiparticles by modifying magnetic confinement, Landau quantization, wave-function localization, and electronic spectra.\cite{DeMartino2007,Masir2008,Ghosh2009} Among the few exactly solvable field profiles, the exponentially decaying magnetic field enabled Ghosh to obtain the exact Landau spectrum of untilted massless Dirac fermions.\cite{Ghosh2009} An exact Lorentz-covariant formulation for tilted anisotropic Dirac fermions was subsequently developed by Islam and Jayannavar, albeit only for uniform magnetic fields.\cite{Islam2017}


Besides externally applied magnetic fields, mechanical strain can generate spatially varying gauge fields and pseudo-magnetic fields capable of producing Landau quantization without external magnetic fields.\cite{Pereira2009,Guinea2010,Levy2010,Vozmediano2010} While the exponentially decaying magnetic-field problem is exactly solvable for untilted Dirac fermions, and the Lorentz approach has been established for tilted anisotropic Dirac fermions in a uniform magnetic field, their combination for an exponentially decaying field has not been addressed. The resulting inverse Lorentz transformation is nontrivial because it renders the effective guiding-center parameter energy dependent. Such a theory is important because it combines the Lorentz-boost physics of tilted Dirac cones with the spatially varying magnetic confinement of a NUMF.
%

In this work, we combine the exact solution for Dirac fermions in an exponentially decaying magnetic field with the Lorentz mapping of a tilted anisotropic Dirac cone. After removing the anisotropy by coordinate rescaling, a Lorentz boost eliminates the intrinsic tilt and maps the problem onto the exactly solvable isotropic Dirac Hamiltonian in the exponentially decaying magnetic field\cite{Ghosh2009}. Upon transforming the resulting spectrum back to the laboratory frame, the boosted guiding-center momentum becomes coupled to the laboratory-frame energy, leading to an intrinsically energy-dependent guiding-center parameter and an implicit Landau-level quantization condition. This provides the central extension of the existing uniform-field Lorentz treatment to the exponentially decaying magnetic-field profile considered here.

\textbf{\textit{Continuum Hamiltonian}:}
We consider the low-energy quasiparticles in a tilted anisotropic Dirac material
subjected to an exponentially decaying magnetic field, \(
\mathbf{B}(\textbf{r})=B_0e^{-\lambda x}\hat{\mathbf{z}}
\). We choose the Landau gauge specified by
\(
	\mathbf{A}(\textbf{r})
	=
	\left(
	0,
	-{B_0}/{\lambda}e^{-\lambda x},
	0
	\right),
	\label{eq:Gauge}
\) such that the effective
single-particle Hamiltonian becomes
\begin{equation}
	H
	=
	v_x\sigma_xp_x
	+
	v_y\sigma_y(p_y+e A_y(x))
	+
	v_t(p_y+e A_y(x)) \mathcal{I}.
	\label{eq:Hamiltonian}
\end{equation}
Here $\sigma_i$ are the Pauli matrices and \(\mathcal{I}\) is the identity matrix. 
The trial 
wavefunction for the Hamiltonian in Eq. \eqref{eq:Hamiltonian} is therefore written as
\(
	\Psi(x,y)
	=
	e^{ik_y y}
	\begin{pmatrix}
		\psi_A(x)&
		\psi_B(x)
	\end{pmatrix}^\mathcal{T},
\)
with conserved momentum $k_y$. Substituting it
into the Schroedinger equation, $H\Psi=\mathcal{E}\Psi$ gives
\begin{equation}
	\begin{aligned}
		&\left[\mathcal{E}-v_t\left(p_y+eA_y(x)\right)\right]\psi_{A/B}
		\\
		&\qquad =
		\left[
		v_xp_x
		\mp iv_y\left(p_y+eA_y(x)\right)
		\right]\psi_{B/A},
	\end{aligned}
	\label{eq:coupled}
\end{equation}
To remove the velocity anisotropy, we introduce the rescaled coordinates
\(\tilde{x}
	=
	\sqrt{{v_y}/{v_x}}\,x,
	\;
	\tilde{y}
	=
	\sqrt{{v_x}/{v_y}}\,y,
\)
together with the corresponding canonical momenta
\(\tilde{p}_x
	=
	\sqrt{{v_x}/{v_y}}\,p_x,
	\;
	\tilde{p}_y
	=
	\sqrt{{v_y}/{v_x}}\,p_y.
	\)
These transformations preserve the canonical commutation relations,
\(
	[\tilde{x}_i,\tilde{p}_j]
	=
	i\hbar\delta_{ij}.
\)
Together, the vector potential is rescaled consistently as
\(
	\tilde{A}_y(\tilde{x})
	=
	\sqrt{{v_y}/{v_x}}\,
	A_y(x),\)
such that
\begin{equation}
	\tilde{p}_y
	+
	e\tilde{A}_y(\tilde{x})
	=
	\sqrt{\frac{v_y}{v_x}}\,
	\left[
	p_y+eA_y(x)
	\right].
	\label{eq:minimal_coupling_rescaling}
\end{equation}
Introducing the geometric-mean velocity
\(
	v_c=\sqrt{v_xv_y},
\)
the two anisotropic terms transform according to
\begin{equation}
	\begin{aligned}
		v_x p_x
		&=
		v_c\tilde{p}_x,
		\text{and }
		v_y\left[p_y+eA_y(x)\right]
		=
		v_c
		\left[
		\tilde{p}_y
		+
		e\tilde{A}_y(\tilde{x})
		\right],
	\end{aligned}
\end{equation}
with the anisotropic Dirac Hamiltonian consequently becoming
\begin{equation}
	\begin{aligned}
		H=&\tilde{v}_t(\tilde{p}_y+e \tilde{A}_y(\tilde{x})) \mathcal{I}\\
		&+
		v_c\left\{
		\sigma_x\tilde{p}_x
		+\sigma_y
		\left[\tilde{p}_y+e\tilde{A}_y(\tilde{x})\right]
		\right\}.
	\end{aligned}
	\label{eq:isotropic_dirac}
\end{equation}
Here we have defined \(\tilde{v}_t=v_t\sqrt{v_x/v_y}\).
The magnetic field associated with the rescaled vector potential is
\begin{align}
	\tilde{B}_z(\tilde{x})
	&=
	\frac{\partial\tilde{A}_y(\tilde{x})}
	{\partial\tilde{x}}
=
	\sqrt{\frac{v_y}{v_x}}\,
	\frac{\partial A_y(x)}{\partial x}
	\frac{\partial x}{\partial\tilde{x}}
=
	B_z(x).
	\label{eq:magnetic_field_rescaling}
\end{align}
Therefore, the field amplitude is preserved, while its spatial
dependence is expressed in terms of the rescaled coordinate.
For an exponentially decaying magnetic field,
\(	B_z(x)=B_0e^{-\lambda x},
\)
the transformed field takes the form
\begin{equation}
	\tilde{B}_z(\tilde{x})
	=
	B_0
	\exp\left(
	-\lambda
	\sqrt{\frac{v_x}{v_y}}\,
	\tilde{x}
	\right)
	=
	B_0e^{-\tilde{\lambda}\tilde{x}},
\end{equation}
where
\(
	\tilde{\lambda}
	=
	\lambda\sqrt{{v_x}/{v_y}}.
\)
Following the anisotropic rescaling, the Dirac sector is characterized
by the single velocity \(v_c\), which plays the role of
the invariant velocity of the emergent Lorentz symmetry. 
For notational simplicity, from this point onward we suppress the tildes on
the rescaled quantities and write
\begin{equation}
v_t \equiv \tilde{v}_t,
\;
x \equiv \tilde{x},
\;
p_y \equiv \tilde{p}_y,
\;
A_y(x) \equiv \tilde{A}_y(\tilde{x}),
\;
\lambda \equiv \tilde{\lambda},
\end{equation}
with all quantities henceforth understood to refer to the rescaled isotropic
representation unless stated otherwise.
 The tilt-dependent
term can be separated as
\begin{equation}
	\mathcal{E}
	-v_t\left[p_y+eA_y(x)\right]
	=
	\left(\mathcal{E}-v_t p_y\right)
	-ev_t A_y(x).
	\label{eq:tilt_separation}
\end{equation}
This separation distinguishes the part involving the conserved energy
and momentum from the position-dependent contribution arising
from the vector potential.
The latter can be identified with an effective scalar-potential term, and
accordingly, we define
\(
	\phi_{\mathrm{eff}}(x)
	=
v_t A_y(x).
	\label{eq:effective_scalar_potential}
\)
The tilt-dependent combination can therefore be written as
\begin{equation}
	\mathcal{E}
	-v_t\left[p_y+eA_y(x)\right]
	=
	\mathcal{E}
	-v_t p_y
	-e\phi_{\mathrm{eff}}(x).
	\label{eq:tilt_effective_potential}
\end{equation}
For the exponentially decaying vector potential
\(	A_y(x)
	=
	-{B_0}/{\lambda}e^{-\lambda x},
\)
the effective scalar potential becomes
\begin{equation}
	\phi_{\mathrm{eff}}(x)
	=
	-\frac{v_t B_0}{\lambda}
	e^{-\lambda x}.
	\label{eq:effective_scalar_exp}
\end{equation}
Unlike the uniform-field problem, where the effective scalar potential is
linear in the spatial coordinate owing to the constant electric field, the
present exponentially decaying magnetic field generates a spatially varying
effective electric field. Consequently, the associated effective scalar
potential,
\(
\phi_{\rm eff}(x)=v_tA_y(x),
\)
also acquires the same exponential profile as the vector potential. Since
\(E_{\rm eff}(x)=-d\phi_{\rm eff}/dx\), there is no requirement for the
scalar potential to be linear in position. The resulting \(\phi_{\rm eff}(x)\)
therefore constitutes the nonuniform counterpart of the effective scalar
potential appearing in the corresponding uniform-field problem.

\textbf{\textit{Lorentz-Covariant Mapping of the Tilted Anisotropic Dirac Hamiltonian}:}
The corresponding effective electric field is
\(
\mathbf{E}_{\mathrm{eff}}
=
-\boldsymbol{\nabla}\phi_{\mathrm{eff}}.
\)
Since the potential depends only on \(x\), we obtain
\begin{equation}
E_{\mathrm{eff},x}(x)
=
-v_t\frac{dA_y(x)}{dx}
=
-v_t B_z(x).
\label{effect_E}
\end{equation}
which vectorially is written as
\(\mathbf{E}_{\mathrm{eff}}
	=
	-v_t B_z(x)\,\hat{\mathbf{x}}.
\)
The sign is determined by our gauge and charge conventions. For the
exponentially decaying magnetic field,
\(
B_z(x)=B_0e^{-\lambda x},
\)
Eq.~\eqref{effect_E} gives
\(
E_{\mathrm{eff},x}(x)=-v_tB_0e^{-\lambda x}.
\)
Thus, the effective electric and magnetic fields share the same spatial dependence, with
\(
{E_{\mathrm{eff},x}(x)}/{B_z(x)}=-v_t,
\)
independent of \(x\). This follows from the proportionality
\(\phi_{\mathrm{eff}}(x)=v_tA_y(x)\), which permits a single global Lorentz boost with \(u=v_t\). This construction is specific to the proportional scalar and vector potentials and does not generally apply to arbitrary nonuniform crossed fields.

\noindent\textit {Transformation of the Electric Field}: Consider a boost along the \(y\)-direction,
\(
\mathbf{u}
=
u\,\hat{\mathbf{y}},
\)
and define
\(
\beta_u
=
{u}/{v_c},
\;
\gamma_u
={1}/{\sqrt{1-\beta_u^2}}.
\)
For mutually perpendicular electric and magnetic fields, the transverse
electric field transforms as
\(
	\mathbf{E}_{\mathrm{eff}}^{\,\prime}
	=
	\gamma_u
	\left(
	\mathbf{E}_{\mathrm{eff}}
	+
	\mathbf{u}\times\mathbf{B}
	\right).
\)
In the present case,
\(
\mathbf{u}\times\mathbf{B}
=
u B_z(x)\,\hat{\mathbf{x}}.
\)
Therefore,
\begin{eqnarray}
	\mathbf{E}_{\mathrm{eff}}^{\,\prime}
	=
	\gamma_u
	\left(u-v_t\right)
	B_z(x)\,\hat{\mathbf{x}}.\label{E_lorentz}
\end{eqnarray}
The required boost now follows immediately. To eliminate the effective
electric field in the Lorentz frame, we impose
\(
\mathbf{E}_{\mathrm{eff}}^{\,\prime}=0.
\)
Since
\(
B_z(x)\neq 0,
\)
Eq.~\eqref{E_lorentz} requires
\(	u=v_t,
\)
which is the unique boost velocity that eliminates the effective
electric field in the Lorentz frame.

\noindent\textit {Transformation of the Magnetic Field}:
For mutually perpendicular electric and magnetic fields, the magnetic
field transforms according to
\begin{equation}
	\mathbf{B}_{\perp}^{\,\prime}
	=
	\gamma_u
	\left(
	\mathbf{B}_{\perp}
	-
	\frac{1}{v_c^2}
	\mathbf{u}\times\mathbf{E}_{\mathrm{eff}}
	\right).
\end{equation}
For the present fields,
\(
\mathbf{u}\times\mathbf{E}_{\mathrm{eff}}
=
u\hat{\mathbf{y}}
\times
\left[
-v_t B_z(x)\hat{\mathbf{x}}\right]=
u v_t B_z(x)\hat{\mathbf{z}}
.
\)
Therefore,
\begin{equation}
	B_z^{\prime}
	=
	\gamma_u
	\left(
	1-\frac{u v_t}{v_c^2}
	\right)
	B_z(x).
\end{equation}
Now imposing
\(
u=v_t,
\)
we find
\(
B_z^{\prime}
=
\gamma
\left(
1-{v_t^2}/{v_c^2}
\right)
B_z(x)={B_z}/{\gamma}.
\) We define \(\gamma
={1}/{\sqrt{1-\beta^2}}\) with \(\beta=v_t/v_c\).
Because the boost is along the \(y\)-direction,
\(
x^\prime=x,
\)
and consequently, the exponentially decaying magnetic field
\(
B_z(x)
=
B_0e^{-\lambda x}
\)
transforms as
\begin{equation}
	B_z^{\prime}(x)
	=
	\frac{B_0}{\gamma}
	e^{-\lambda x}.
\end{equation}
Thus, the boosted problem again contains a purely exponentially
decaying magnetic field, with the field magnitute
\(
	B_0^{\prime}
	=
{B_0}/{\gamma},
	\text{ and }	\lambda^{\prime}
	=
	\lambda.
\)

\noindent\textit{Lorentz transformation}:
It is natural to introduce separately the 
energy-momentum vector and the effective electromagnetic potential defined as
\begin{equation}
	p^\mu
	=
	\begin{pmatrix}
		\mathcal{E}/v_c \\[2pt]
		p_x \\[2pt]
		p_y
	\end{pmatrix},
\;	A_{\mathrm{eff}}^\mu
	=
	\begin{pmatrix}
		\phi_{\mathrm{eff}}/v_c \\[2pt]
		A_x \\[2pt]
		A_y
	\end{pmatrix}.
	\label{eq:effective_four_potential}
\end{equation}
The former determines the transformation of the energy and
momentum, whereas the latter describes the transformation of the
effective electromagnetic background.
For the Lorentz boost discussed above, 
the corresponding Lorentz matrix in \((\textbf{2+1})\) dimensions is
\begin{equation}
	\Lambda^\mu{}_{\nu}
	=
	\begin{pmatrix}
		\gamma & 0 & -\beta\gamma \\[2pt]
		0      & 1 & 0 \\[2pt]
		-\beta\gamma & 0 & \gamma
	\end{pmatrix}.
	\label{eq:Lorentz_matrix}
\end{equation}
The canonical energy-momentum vector and the effective electromagnetic
potential transform independently according to
\(
	{p^\prime}^{\mu}
	=
	\Lambda^\mu{}_{\nu}p^\nu,
	\;
	A_{\mathrm{eff}}^\prime{}^{\mu}
	=
	\Lambda^\mu{}_{\nu}A_{\mathrm{eff}}^\nu.
	\label{eq:Lorentz_transform_vectors}
\)
The transformation of the canonical energy-momentum vector leads to
\begin{align}
	\mathcal{E}^\prime
	&=
	\gamma
	\left(
	\mathcal{E}-v_t p_y
	\right),
	\label{eq:energy_transform}
	\\
	p_y^\prime
	&=
	\gamma
	\left(
	p_y-\frac{v_t}{v_c^2}\mathcal{E}
	\right),
	\label{eq:py_transform}\\
	p_x^\prime&=p_x.
\end{align} Since \(p_y=\hbar k_y\) and
\(p_y^\prime=\hbar k_y^\prime\), Eq.~\eqref{eq:py_transform} immediately yields
\begin{equation}
	k_y^\prime
	=
	\gamma
	\left(
	k_y-\frac{v_t}{\hbar v_c^2}\mathcal{E}
	\right)
	=
	\gamma
	\left(
	k_y-\beta\frac{\mathcal{E}}{\hbar v_c}
	\right).
	\label{eq:ky_transform}
\end{equation}
The effective electromagnetic potential transformation in Eq.~\eqref{eq:effective_four_potential}, simultaneously leads to 
\begin{align}
	\frac{\phi_{\mathrm{eff}}^\prime}{v_c}
	&=
	\gamma
	\left(
	\frac{\phi_{\mathrm{eff}}}{v_c}
	-\beta A_y
	\right),
	\label{eq:phi_eff_transform}
	\\
	A_y^\prime
	&=
	\gamma
	\left(
	A_y-\beta\frac{\phi_{\mathrm{eff}}}{v_c}
	\right).
	\label{eq:Ay_transform}
\end{align}
Using \(\phi_{\mathrm{eff}}=v_tA_y=\beta v_cA_y\), Eqs.~(23) and (24) give
\[
\phi_{\mathrm{eff}}^\prime=0,\qquad
A_y^\prime=\frac{A_y}{\gamma}.
\]
Thus, the Lorentz boost with velocity \(v_t\) eliminates only the
tilt-induced effective scalar potential, while the magnetic vector
potential remains finite and is simply rescaled. Consequently,
\(B_z^\prime(x)=B_z(x)/\gamma\), and the boosted Hamiltonian retains the
minimal-coupling combination \(p_y^\prime+eA_y^\prime(x)\), describing a purely
magnetic Dirac problem in the comoving frame.

Having transformed the canonical variables and the electromagnetic
	potential separately, we now write the boosted Dirac Hamiltonian directly
	in terms of the minimally coupled canonical variables. Since the effective
	scalar potential vanishes in the comoving frame, the boosted Hamiltonian
	takes the standard untilted isotropic Dirac form
	\begin{equation}
		H^\prime
		=
		v_c
		\left[
		\sigma_x p_x
		+
		\sigma_y
		\left(
		p_y^\prime+eA_y^\prime(x)
		\right)
		\right].
		\label{eq:boosted_hamiltonian}
	\end{equation}
Equation~(\ref{eq:boosted_hamiltonian}) constitutes the central result of the Lorentz-covariant
mapping. The combined anisotropy rescaling and Lorentz boost eliminate
both the velocity anisotropy and the intrinsic tilt,
\(
H(v_x,v_y,v_t)
\rightarrow
H(v_c,v_c,0),
\)
leaving a purely magnetic isotropic Dirac Hamiltonian. The only effect
of the boost is the reduction of the magnetic-field amplitude,
$B_0^\prime=B_0/\gamma$, while the exponential profile remains unchanged.
Consequently, the boosted Hamiltonian is identical to the exactly
solvable Dirac problem in an exponentially decaying magnetic field,
allowing the Landau spectrum to be obtained analytically. The effects
of tilt and anisotropy are restored only after transforming back to
the laboratory frame.
	
\textbf{\textit{Landau Quantization in the Lorentz Frame}:}	The two-component Dirac spinor transforms under the corresponding spinor representation of the Lorentz boost. Since the boosted Hamiltonian can be constructed directly from the transformed canonical variables and electromagnetic potentials discussed earlier, an explicit representation of the spinor boost operator is not required here.
 Writing \(
		\Psi^\prime
		=
		\begin{pmatrix}
			\psi_A^\prime&
			\psi_B^\prime
		\end{pmatrix}^\mathcal{T},
\)
	the eigenvalue equation
\(
		H^\prime\Psi^\prime=\mathcal{E}^\prime\Psi^\prime
\)
	yields the coupled equations
	\begin{equation}
		\begin{aligned}
			\mathcal{E}^\prime\psi_{A/B}^\prime
			&=
			v_c
			\left[
			p_x
			\mp
			i\left(
			p_y^\prime+eA_y^\prime(x)
			\right)
			\right]\psi_{B/A}^\prime.
		\end{aligned}
		\label{eq:boosted_coupled}
	\end{equation}
The two spinor components can now be decoupled by successively applying
	the corresponding first-order operators. This gives
\begin{equation}
	\begin{aligned}
		{\mathcal{E}^\prime}^{\,2}\psi_{A/B}^\prime
		={}&v_c^2
		\left[p_x\mp i(p_y^\prime+eA_y^\prime)\right]
		\\
		&\times
		\left[p_x\pm i(p_y^\prime+eA_y^\prime)\right]
		\psi_{A/B}^\prime .
	\end{aligned}
	\label{eq:boosted_decoupling}
\end{equation}
Using the operator identity
	\begin{align}
		&
		\left[
		p_x
		\mp
		i\left(
		p_y^\prime+eA_y^\prime(x)
		\right)
		\right]
		\left[
		p_x
		\pm
		i\left(
		p_y^\prime+eA_y^\prime(x)
		\right)
		\right]
		\nonumber\\
		& =
		p_x^{\,2}
		+
		\left(
		p_y^\prime+eA_y^\prime(x)
		\right)^2
		\pm
		i
		\left[
		p_x,
		p_y^\prime+eA_y^\prime(x)
		\right],
		\label{eq:operator_identity}
	\end{align}
	and noting that
	\(	[p_x,p_y^\prime]=0,
	\)
	we obtain
	\begin{align}
		\left[
		p_x,
		p_y^\prime+eA_y^\prime(x)
		\right]
		&=
		e[p_x,A_y^\prime(x)]
		=
		-i\hbar e
		\frac{dA_y^\prime(x)}{dx}
		\nonumber\\
		&=
		-i\hbar eB_z^\prime(x).
		\label{eq:boosted_commutator}
	\end{align}
	Consequently, the decoupled equations reduce to
	\begin{equation}
		\left[
		p_x^{\,2}
		+
		\left(
		p_y^\prime+eA_y^\prime(x)
		\right)^2
		\pm
		\hbar eB_z^\prime(x)
		\right]
		\psi_{A/B}^\prime
		=
		\frac{{\mathcal{E}^\prime}^{\,2}}{v_c^2}
		\psi_{A/B}^\prime.
		\label{eq:decoupled_boosted}
\end{equation}
Here the upper sign corresponds to the $A$ component and the lower sign
to the $B$ component. Equation~(\ref{eq:decoupled_boosted}) is the
second-order equation governing the spinor components in the Lorentz
frame.
For the exponentially decaying magnetic field in the Lorentz frame, the corresponding magnetic length is
\(
	\ell_{B^\prime}
	=
	\sqrt{{\hbar}/{eB_0^\prime}}
	=
	\sqrt{\gamma}\,\ell_B,
	\text{ with }
	\ell_B
	=
	\sqrt{{\hbar}/{eB_0}}.
\)
We now introduce the dimensionless variable
\(\xi
		=
		{e^{-\lambda x}}/
		{(\lambda\ell_{B^\prime})^2}.
\)
With these definitions, Eq.~\eqref{eq:decoupled_boosted} reduces to
\begin{equation}
		\frac{d^2\psi_{A/B}^\prime}{d\xi^2}
		+
		\frac{1}{\xi}
		\frac{d\psi_{A/B}^\prime}{d\xi}
		+
		\left[-
		\frac{\beta_G^2}{\xi^2}
		+
		\frac{2\xi^\prime_0\mp1}{\xi}
		-
		1
		\right]
		\psi_{A/B}^\prime
		=
		0,
	\label{eq:Whittaker_form}
\end{equation}
where
\(
		\beta_G^2
		=-
		\frac{{\mathcal{E}^\prime}^{\,2}}
		{\hbar^2v_c^2\lambda^2}
		+
		(\xi^\prime_0)^2.
\)
Thus, the original tilted anisotropic Dirac problem is transformed,
through the anisotropy rescaling and the Lorentz boost, into a
second-order differential equation of the same form as that for an
untilted Dirac fermion in an exponentially decaying magnetic field.
\cite{Ghosh2009}

To obtain normalizable solutions, we introduce the ansatz
\(
	\psi(\xi)
	=
	\xi^{\beta_G}
	e^{-\xi}
	F(\xi).
\)
Substituting this form into Eq.~\eqref{eq:Whittaker_form} transforms the
differential equation into the standard Kummer (confluent
hypergeometric) equation,
\begin{equation}
	zF_{zz}
	+
	(c-z)F_z
	-
	a_KF
	=
	0,
\end{equation}
where
\(z=2\xi,
\)
with
\(
	c
	=
	2\beta_G+1,
\)
and
\begin{equation}\label{eq:Kummer}
	a_K
	=
	\beta_G+1-\xi^\prime_0
	\pm\frac12.
\end{equation}
The corresponding wavefunction is therefore
\begin{equation}
\boxed{
	\psi(\xi)
	=
	\xi^{\beta_G}e^{-\xi}
	\left[
	C_1\,{}_1F_1(a_K,c,2\xi)
	+
	C_2\,U(a_K,c,2\xi)
	\right],
}
\label{eq:WaveExact}
\end{equation}
where $C_1$ and $C_2$ are arbitrary integration constants.
Regularity at $\xi=0$ together with normalizability as
$\xi\rightarrow\infty$ requires the confluent hypergeometric series to
terminate, giving the polynomial condition
\(
	a_K=-n,
\text{ with }
	n=0,1,2,\cdots.
\)
Using Eq.~\eqref{eq:Kummer}, one obtains
\(
\beta_G=\xi^\prime_0-n.
\)
For the finite-energy bound states, normalizability of the asymptotic factor
$\xi^{\beta_G}$ requires $\beta_G>0$, and hence
\begin{equation}
n<\xi^\prime_0=\frac{|k^\prime_y|}{\lambda}.
\label{eq:allowed_n}
\end{equation}
Thus, for a fixed $k^\prime_y$, only a finite number of Landau levels are supported by the exponentially decaying magnetic field. Subject to this restriction, the exact boosted-frame spectrum is
\begin{equation}
{\mathcal{E}^\prime}^\pm_n=\pm\hbar v_c\lambda
\sqrt{(\xi^\prime_0)^2-(\xi^\prime_0-n)^2}.
\end{equation}
The $n=0$ mode is considered separately below from the first-order Dirac equation.

\begin{figure}[t]
	\includegraphics[width=0.235\textwidth,height=3.0cm]{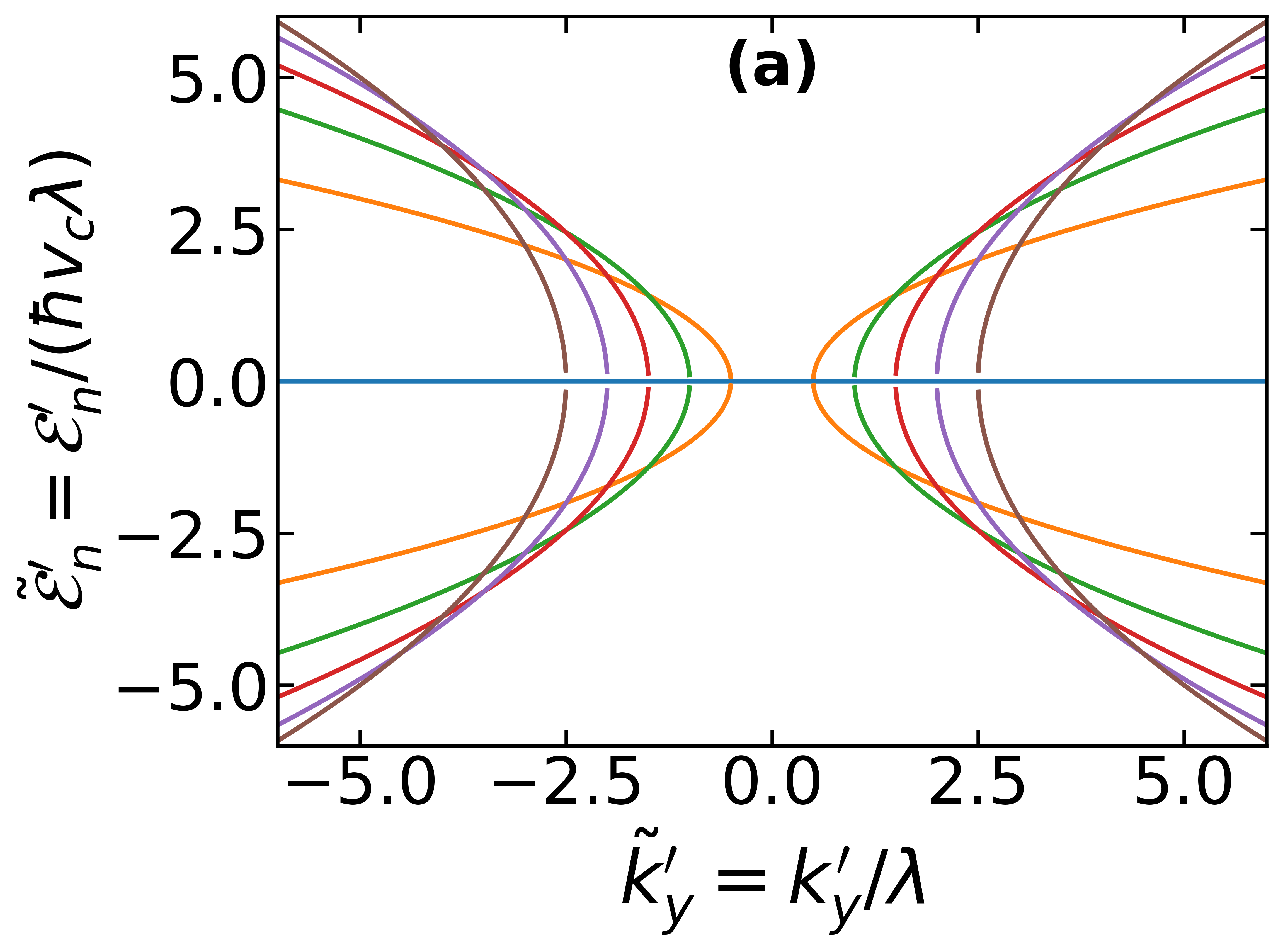}
	\includegraphics[width=0.235\textwidth,height=3.0cm]{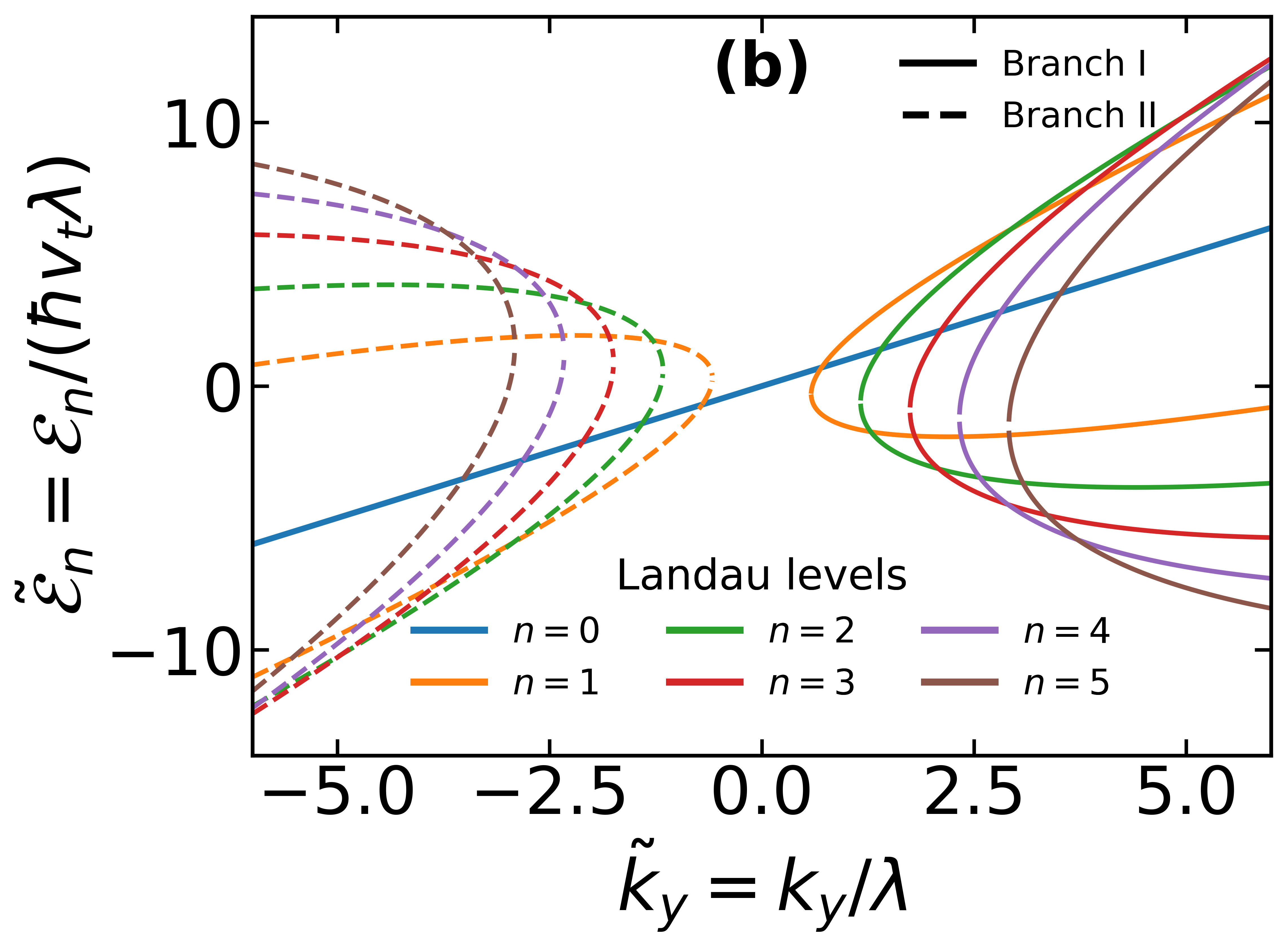}
	\caption{(Color online) (a) Boosted-frame Landau-level spectrum $\tilde{\mathcal{E}}_n^{\prime\,\pm}$ and (b) corresponding laboratory-frame spectrum $\tilde{\mathcal{E}}^\pm_n$. We use $v_x=0.86\times10^6$ m/s, $v_y=0.69\times10^6$ m/s, and $v_t=0.32\times10^6$ m/s. In (b), solid and dot-dashed curves denote \textit{Branches I} and \textit{II}, respectively. Only roots satisfying the branch conditions and $\pm\gamma(\mathcal{E}_n^\pm-v_t\hbar k_y)\geq0$ are retained. The zeroth Landau level is dispersionless in the boosted frame and linearly dispersive in the laboratory frame.
	}
	\label{fig:boost_lab_spectrum}
\end{figure}

\textit{Zeroth Landau level:}
	The zeroth Landau level is most directly obtained from the first-order
	Dirac equations in Eq.~\eqref{eq:boosted_coupled}. For the present field convention, we seek
	a normalizable state with $\psi_A^\prime=0$ and $\psi_B^\prime\neq0$. Eq.~\eqref{eq:boosted_coupled}
	then immediately gives
	\begin{equation}
	\mathcal{E}_0^\prime=0,
	\qquad
	\left[p_x-i\left(p_y^\prime+eA_y^\prime(x)\right)\right]\psi_B^\prime=0 .
	\end{equation}
	The normalized zero mode therefore
	has the form
	\begin{equation}
	\Psi_0^\prime(x,y)=
	\mathcal N_0 e^{ik_y^\prime y}
	\begin{pmatrix}
	0\\[1mm]
	\displaystyle
	\exp\!\left[-k_y^\prime x-
	\frac{e^{-\lambda x}}{(\lambda l_{B^\prime})^2}\right]
	\end{pmatrix},
	\label{eq:zero_mode}
	\end{equation}
	which is normalizable for $k_y^\prime> 0$. For the opposite field direction,
	the surviving spinor component is interchanged. Thus, the zeroth level
	is pinned at $\mathcal{E}_0^\prime=0$ in the boosted frame, whereas the inverse Lorentz
	transformation $\mathcal{E}^\prime=\gamma(\mathcal{E}-v_t\hbar k_y)$ gives
	$\mathcal{E}_0=\hbar v_t k_y$ in the laboratory frame, explaining its linear
	dispersion in Fig.~\ref{fig:boost_lab_spectrum} (a) and \ref{fig:boost_lab_spectrum} (b).

\textbf{\textit{The Laboratory Spectrum}:} Using the Lorentz transformation of the four-momentum,
\(
p_y^\prime
=
\gamma
\left(
p_y
-
{v_t}/{v_c^2}\mathcal{E}_n
\right),
\)
and recalling that
\(
p_y=\hbar k_y,
p_y^\prime=\hbar k_y^\prime,
\)
one obtains
\(	k_y^\prime
	=
	\gamma
	\left(
	k_y
	-
	{v_t\mathcal{E}_n}/{\hbar v_c^2}
	\right).
\)
Hence,
\begin{equation}
	\xi_0^\prime
	=
	\frac{\gamma}{\lambda}
	\left|
	k_y
	-
	\frac{v_t\mathcal{E}_n}{\hbar v_c^2}
	\right|.
\end{equation}
Therefore, although $\xi_0^\prime$ is a constant quantum number in the boosted frame, it becomes an energy-dependent quantity when expressed in terms of the laboratory variables. Consequently, the boosted-frame Landau-level spectrum,
\begin{equation}\label{boost_E}
{\mathcal{E}^\prime}^\pm_n
=\pm
\hbar v_c\lambda
\sqrt{
	2n\xi_0^\prime-n^2
}.
\end{equation}
can be transformed back to the laboratory frame only through the Lorentz transformations of energy and momentum.

The spectrum in Eq.~\eqref{boost_E} has the same functional form as that in Ref. [\cite{Ghosh2009}], reflecting the Lorentz-covariant mapping of the tilted anisotropic Hamiltonian onto an isotropic Dirac problem in the boosted frame. Unlike the untilted case, however, the boosted guiding-center parameter $k_y^\prime$ is no longer an independent quantum number. Through the Lorentz transformation it becomes coupled to the laboratory-frame energy, rendering the guiding-center coordinate intrinsically energy dependent. Consequently, the laboratory-frame spectrum cannot be expressed explicitly in terms of $k_y$ but is instead determined by a self-consistent implicit quantization condition. This energy-dependent guiding center is a direct consequence of the interplay between Dirac-cone tilt and magnetic-field inhomogeneity.

Since $\xi_0^\prime$ depends on the conserved momentum $k_y^\prime$ in the boosted
frame, substituting
\(
	k_y^\prime
	=
	\gamma
	\left(
	k_y-{\beta \mathcal{E}_n}/{\hbar v_c}
	\right)
\)
into the boosted-frame spectrum yields
\begin{equation}
	{\mathcal{E}^\prime}^\pm_n
	=\pm
	\hbar v_c
	\sqrt{
		2n\gamma\lambda
		\left|
		k_y-\frac{\beta \mathcal{E}_n}{\hbar v_c}
		\right|
		-
		n^2\lambda^2
	}.
\end{equation}

To obtain the spectrum in the laboratory frame, we employ the inverse
Lorentz transformation discussed earlier, such that 
\begin{equation}
	\mathcal{E}^\pm_n-v_t\hbar k_y
	=\pm
	\frac{\hbar v_c}{\gamma}
	\sqrt{
		2n\gamma\lambda
		\left|
		k_y-\frac{\beta \mathcal{E}_n}{\hbar v_c}
		\right|
		-
		n^2\lambda^2
	}.
\end{equation}
To express the spectrum fully in terms of the original laboratory-frame
variables, we finally undo the anisotropic rescaling introduced earlier.
Restoring the original variables according to
\(
{k}_y
\rightarrow
\sqrt{{v_y}/{v_x}}\,k_y,
\;
{v}_t
\rightarrow
v_t\sqrt{{v_x}/{v_y}},
\;
{\lambda}
\rightarrow
\lambda\sqrt{{v_x}/{v_y}},
\)
while the energy remains unchanged.
Furthermore, on using
\(
v_c=\sqrt{v_xv_y},
\) 
\(\beta
=
{{v}_t}/{v_c}
\rightarrow
{v_t}/{v_y},
\;
\gamma
=
{1}/{\sqrt{1-\beta^2}},
\)
the Landau-level spectrum can be written entirely in terms of the
original anisotropic laboratory-frame variables as
\begin{equation}
	\mathcal{E}^\pm_n-v_t\hbar k_y
	=\pm
	\frac{\hbar v_c}{\gamma}
	\sqrt{
		2n\gamma\lambda
		\left|
		k_y-\frac{\beta\mathcal{E}_n}{\hbar v_y}
		\right|
		-
		n^2\lambda^2\frac{v_x}{v_y}
	}
.
\label{eq:physical_lab_spectrum}
\end{equation}
Equation~(\ref{eq:physical_lab_spectrum}) therefore represents the
spectrum in the physical laboratory frame, with the anisotropic
rescaling fully reversed.

Before proceeding further, it is useful to examine the origin of the laboratory-frame spectrum. In the boosted frame, the Landau levels form a single symmetric spectrum because the Lorentz transformation removes the tilt. Transforming back to the laboratory frame couples the conserved momentum $k_y^\prime$ to the energy, so that the absolute-value dependence of the guiding-center parameter gives rise to two solution branches corresponding to the two signs of the Lorentz-transformed momentum. These branches represent the two admissible laboratory-frame realizations of the same Lorentz-covariant spectrum.

Accordingly, squaring Eq.~\eqref{eq:physical_lab_spectrum} gives the implicit quantization condition in the laboratory frame,
\begin{equation}
	\left(
	\mathcal{E}^\pm_n-v_t\hbar k_y
	\right)^2
	=
	\frac{2n\hbar^2v_c^2\lambda}{\gamma}
	\left|
	k_y-\frac{\beta\mathcal{E}^\pm_n}{\hbar v_y}
	\right|
	-
	\frac{\hbar^2v_c^2\lambda^2n^2}{\gamma^2}
	\frac{v_x}{v_y}
.
\end{equation}
Introducing
\(
S
=
k_y-{\beta \mathcal{E}_n}/{\hbar v_y},
\)
the absolute value separates the spectrum naturally into two branches.

\noindent\textit{Branch I: $S\ge0$, or \( |S|=S\)}:
This branch corresponds to positive values of the Lorentz-transformed guiding-center parameter. The implicit quantization condition therefore reduces to
\begin{eqnarray}\label{lab_E_I}
\left(\tilde{\mathcal{E}}^\pm_n
-
\tilde{k}_y\right)^2
=
\frac{2n}{\gamma \beta^2}
\left(
\tilde{k}_y-\beta^2\tilde{\mathcal{E}}^\pm_n
\right)
-
\frac{1}{\beta^2}\frac{n^2}{\gamma^2}\frac{v_x}{v_y}
,
\end{eqnarray}
after introducing the dimensionless variables
\(
\tilde{\mathcal{E}}^\pm_n=\mathcal{E}^\pm_n/(\hbar v_t\lambda)
\)
and
\(
\tilde{k}_y=k_y/\lambda.
\)

\noindent{\textit{Branch II: $S\le0$, or \(|S|=-S\)}}:
For negative values of the Lorentz-transformed guiding-center parameter, the absolute value changes sign, yielding
\begin{eqnarray}\label{lab_E_II}
\left(\tilde{\mathcal{E}}^\pm_n
-
\tilde{k}_y\right)^2
=-
\frac{2n}{\gamma \beta^2}
\left(
\tilde{k}_y-\beta^2\tilde{\mathcal{E}}^\pm_n
\right)
-
\frac{1}{\beta^2}\frac{n^2}{\gamma^2}\frac{v_x}{v_y}
.
\end{eqnarray}
The two branches are therefore distinguished solely by the sign of the Lorentz-transformed guiding-center parameter and together constitute the complete laboratory-frame Landau-level spectrum.

Figure~\ref{fig:boost_lab_spectrum} illustrates the evolution of the Landau-level spectrum under the inverse Lorentz transformation. In the boosted frame [Fig.~\ref{fig:boost_lab_spectrum}(a)], the spectrum consists of symmetric electron and hole branches described by the analytical solution [Eq.~\eqref{boost_E}], while the zeroth Landau level remains pinned at $\tilde{\mathcal{E}}_0^\prime=0$ (non-dispersive). As the Landau-level index increases, the spectrum extends to larger values of $|\tilde{k}_y^\prime|$, reflecting the increasing guiding-center coordinate in the exponentially decaying magnetic field.
The corresponding laboratory-frame spectrum is shown in Fig.~\ref{fig:boost_lab_spectrum}(b). The inverse Lorentz transformation renders the spectrum asymmetric, producing two solution branches described by Eqs.~(\ref{lab_E_I}) and (\ref{lab_E_II}). Near $\tilde{k}_y=0$, the branches connect smoothly, whereas for larger $|\tilde{k}_y|$ they evolve into distinct positive- and negative-energy dispersions. The solid and dot-dashed curves denote Branches~I and II, respectively.

The quadratic equations for the two branches admit closed-form solutions. Since they are obtained by squaring the original eigenvalue equation, however, not every algebraic root corresponds to a physical Landau level. The admissible spectrum is determined by enforcing both the branch conditions and the original eigenvalue equation. In particular, the requirement
\(\pm	\gamma\left(\mathcal{E}_n-v_t\hbar k_y\right)\ge0\)
is not preserved under squaring and removes the spurious roots introduced by the quadratic equations. The complete laboratory-frame spectrum is therefore obtained from the union of Branches~I and II after imposing these admissibility conditions. The emergence of the two-branch structure is a direct consequence of the inverse Lorentz transformation and has no analogue in the symmetric boosted-frame spectrum.




\textbf{\textit{Asymptotic Expansion in the Uniform-Field Limit}:} The parameter $\lambda^{-1}$ represents the characteristic magnetic
decay length of the exponentially decaying field
$B(x)=B_0e^{-\lambda x}$.
Consequently, the limit $\lambda\rightarrow0$ continuously transforms
the nonuniform magnetic field into a spatially uniform one.
The analysis below demonstrates that the exact boosted-frame solution
correctly reproduces the well-known relativistic Landau spectrum in
this limit. 
The origin of the momentum scaling in the uniform-field limit is most transparently understood from the vector potential. In the boosted frame,
	\begin{equation}
	A_y^\prime(x)=-\frac{B_0^\prime}{\lambda}e^{-\lambda x}
	=-\frac{B_0^\prime}{\lambda}+B_0^\prime x+O(\lambda),
	\; \lambda\rightarrow0 .
\end{equation}
The divergent constant $-B_0^\prime/\lambda$ is gauge dependent and does not contribute to the magnetic field. Since the Hamiltonian depends on the gauge-invariant combination $\hbar k_y^\prime+eA_y^\prime$, this constant is compensated by a corresponding shift of the canonical momentum,
\begin{equation}
	k_y^\prime=\frac{eB_0^\prime}{\hbar\lambda}+k_0^\prime+O(\lambda)
	=\frac{1}{\lambda l_{B^\prime}^{,2}}+k_0^\prime+O(\lambda),
	\label{eq:ky_uniform}
	\end{equation}
	where $l_{B^\prime}=\sqrt{\hbar/(eB_0^\prime)}$. Consequently,
	\(
	\xi_0^\prime={|k_y^\prime|}/{\lambda}
	=\frac{1}{\lambda^2l_{B^\prime}^{,2}}
	+\frac{k_0^\prime}{\lambda}+O(1),
	\)
	so that $\xi_0^\prime=O(\lambda^{-2})$. Substitution into Eq.~\eqref{boost_E} gives
	\begin{equation}
	{\mathcal{E}^\prime}^\pm_n
	=\hbar v_c
	\sqrt{\frac{2n}{l_{B^\prime}^{,2}}
		+2nk_0^\prime\lambda-n^2\lambda^2+\cdots},
	\end{equation}
	and hence
	\(
	{\mathcal{E}^\prime}^\pm_n\xrightarrow{\lambda\to0}
	\pm{\hbar v_c}/{l_{B^\prime}}\sqrt{2n},
	\)
	recovering the relativistic Landau-level spectrum in a uniform magnetic field as below:
\begin{equation}
\mathcal{E}^\pm_n-v_t\hbar k_y
=
\pm\frac{\hbar v_c}{l_B}
\sqrt{2n}\,
(1-\beta^2)^{3/4}.
\end{equation}
This is precisely the characteristic
\(
(1-\beta^2)^{3/4}
\)
scaling obtained in Ref. [\cite{Islam2017}].

In \textit{\textbf{conclusion}}, we have developed an analytical theory of Landau
quantization for tilted anisotropic Dirac fermions in an exponentially
decaying magnetic field. The laboratory-frame spectrum is described by
an implicit quantization condition originating from the energy-dependent
guiding-center coordinate induced by the inverse Lorentz transformation.
This leads naturally to two admissible branches of Landau levels,
reflecting the nontrivial correspondence between the boosted and
laboratory frames.

The theory continuously reproduces the uniform-field limit, recovering
the known Lorentz-covariant Landau spectrum of tilted Dirac fermions in
a homogeneous magnetic field. This agreement confirms the consistency
of the present formalism while extending it to spatially nonuniform
magnetic fields.

The present results therefore extend the Lorentz-covariant treatment of tilted anisotropic Dirac fermions from the uniform-field case to an exponentially decaying magnetic field. A distinctive consequence of the nonuniform field is that the inverse Lorentz transformation couples the boosted guiding-center momentum to the laboratory-frame energy, resulting in an energy-dependent guiding-center parameter and an implicit Landau-level spectrum. The approach may provide a useful starting point for investigating other spatially varying field profiles for which the corresponding boosted Dirac problem admits an analytical treatment.



\end{document}